\documentclass[journal = nalefd, manuscript=article]{achemso} %, layout=twocolumn

\usepackage{amsmath}
\usepackage{amssymb}
\usepackage{graphicx}
\usepackage{textcomp}
\usepackage[utf8]{inputenc}
\usepackage{braket}
\title{Robust and accurate electric field sensing with solid state spin ensembles}

\author{Julia Michl}
\email{j.michl@pi3.uni-stuttgart.de}
\author{Jakob Steiner}
\author{Andrej Denisenko}
\affiliation{3. Physikalisches Institut, University of Stuttgart, Pfaffenwaldring 57, 70569 Stuttgart, Germany}
\author{André Bülau}
\affiliation{Hahn-Schickard, Allmandring 9b, 70569 Stuttgart, Germany}
\author{André Zimmermann}
\affiliation{Institut für Mikrointegration, University of Stuttgart, Allmandring 9b, 70569 Stuttgart, Germany}
\author{Kazuo Nakamura}
\affiliation{Application Technology Research Institute, Tokyo Gas Company, Ltd.,
Yokohama, 230-0045 Japan}
\author{Hitoshi Sumiya}
\affiliation{Advanced Materials Laboratory, Sumitomo Electric Industries, Ltd., Itami, 664-0016 Japan}
\author{Shinobu Onoda}
\affiliation{Takasaki Advanced Radiation Research Institute, National Institutes for Quantum and Radiological Science and Technology, Takasaki, 370-1292, Japan}
\author{Philipp Neumann}
\email{philipp@nvision-imaging.com}
\affiliation{3. Physikalisches Institut, University of Stuttgart, Pfaffenwaldring 57, 70569 Stuttgart, Germany}
\author{Junichi Isoya}
\affiliation{Faculty of Pure and Applied Sciences, University of Tsukuba, Tsukuba, 305-8573 Japan}
\author{Jörg Wrachtrup}
\affiliation{3. Physikalisches Institut, University of Stuttgart, Pfaffenwaldring 57, 70569 Stuttgart, Germany}

\date{\today}

\begin{document}

\maketitle

%\twocolumn[
%\begin{@twocolumnfalse}
%\onecolumn
\begin{abstract}
	Electron spins in solids constitute remarkable quantum sensors.
	Individual defect centers in diamond were used to detect individual nuclear spins with nanometer scale resolution, and ensemble magnetometers rival SQUID and vapor cell magnetometers when taking into account room temperature operation and size.
	NV center spins can also detect electric field vectors, despite their weak coupling to electric fields. %even that of an isolated fundamental charge, despite their weak coupling to electric fields.
	Here, we employ ensembles of NV center spins to measure macroscopic AC electric vector fields with high precision.
	We utilize low strain, $^{12}$C enriched diamond to achieve maximum sensitivity and tailor the spin Hamiltonian via proper magnetic field adjustment to map out the AC electric field strength and polarization and arrive at refined electric field coupling constants.
	For high precision measurements we combine classical lock-in detection with aspects from quantum phase estimation for effective suppression of technical noise.
	Eventually, this enables $t^{-1/2}$ uncertainty scaling of the electric field strength over extended averaging periods, enabling us to reach a sensitivity down to $10^{-7}$ V/\textmu m. %We achieve an electric field sensitivity of 0.16 (V/cm)/$\sqrt(\text{Hz})$.
\end{abstract}
%\end{@twocolumnfalse}]

%\twocolumn
%\tableofcontents
%\indent
%PRX looks for papers that:
%\begin{itemize}
%	\item make new fundamental experimental discoveries
%	\item new paradigm or shift
%	\item establish fruitful analogy or connection between different fields
%	\item push established into new direction
%	\item advance state-of-the-art of field
%\end{itemize}
%What do we do:
%\begin{itemize}
%	\item ac electric field sensing with ensembles of NV centers in diamond
%	\item Hamiltonian engineering by magnetic field (robust, only E-field, quantitative, exclude temperature DQ!!!)
%	\item technical noise reduction by classical lock-in detection and quantum phase estimation
%	\item unmatched low strain, isotopically engineered hpht diamond
%	\item exact determination of NV Electric field coupling constants
%	\item exact modeling of NV ODMR spectra including strain
%	\item longterm stability of E-field sensing, implications for voltage/field reference
%\end{itemize}

%\section{Introduction}
Solid state spins have matured to leading contenders in the field of nanoscale sensing \cite{reviewFriedemann}. As spins couple to magnetic fields via their magnetic moment, they naturally lend themselves to magnetometry. Additionally, by coupling to the lattice they have been shown to measure temperature and strain, and hence force or pressure with high sensitivity and exquisite spatial resolution\cite{Laraoui2015}. Intrinsically however, spins only interact with electric fields via the orbital angular momentum of e.g. electrons. Yet, precision sensing of electric fields or charges is a long standing challenge in metrology, tackled by e.g. single electron transistors \cite{Lee2008, Vincent2004, CNeumann2013}, energy level shift measurements of ions \cite{Osterwalder1999} or Rydberg atoms\cite{Sedlacek2012}.
Spins with total angular momentum $J\geq 1$ couple to gradients of electric fields and all spins can be indirectly affected by electric fields via spin orbit coupling.
It would thus be interesting to extend those studies to solid state spin systems for the sake of precision metrology especially with high spatial resolution. Electric field coupling to spins has been observed in a variety of cases\cite{Mims, Dolde2011} and used e.g. for their coherent control. 
Point defects in diamond\cite{Kato2003, Iwasaki2017} and e.g. SiC\cite{Wolfowicz2018} typically show little coupling to electric fields because of the small spin orbit coupling in these materials and the fact that the orbital angular momentum is largely quenched at room temperature. However, the NV center shows a sizeable spin orbit coupling in the excited state and a small fraction %($\lambda LS/\delta E$, where $\delta E$ is the energy difference between HOMO and LUMO of the NV center) 
of that is measured in the electron ground state.  
This shift has been detected before using single NVs and in NV ensembles with strong fields, using ESR and ODMR\cite{Dolde2011, Chen2017, Oort1990}.
In the present work we are concerned with precision measurements of weak AC electric fields using a low strain, isotopically engineered HPHT diamond. 
As the NV center in diamond has a $C_{3v}$ symmetry, the linear Stark shift Hamiltonian is given by\cite{Mims}
\begin{align}
\label{eq:hstark}
\hat{H}_{\text{Stark}}= & R_{15}\left(E_x\left(S_xS_z+S_zS_x\right)+E_y\left(S_yS_z+S_zS_y\right)\right)\nonumber\\
&-R_{2E}\left(E_x\left(S_x^2-S_y^2\right)-E_y\left(S_xS_y+S_yS_x\right)\right)\nonumber\\
&+R_{3D}E_z\left(S_z^2-\frac{1}{3}S\left(S+1\right)\right).
\end{align}
with the axial coupling constant $R_{3D}$ being around 3.5 kHz/V/\textmu m and the transversal coupling $R_{2E}$ being around 170 kHz/V/\textmu m. 
The first term is suppressed by the zero-field splitting and the last term is, due to its small coupling constant $R_{3D}$, not viable for precision measurement. Hence the transversal term, $H_{\text{Stark}, \perp}$ given by $R_{2E}$ is of primary importance for this work.\\

\begin{figure}
\includegraphics[width=\columnwidth]{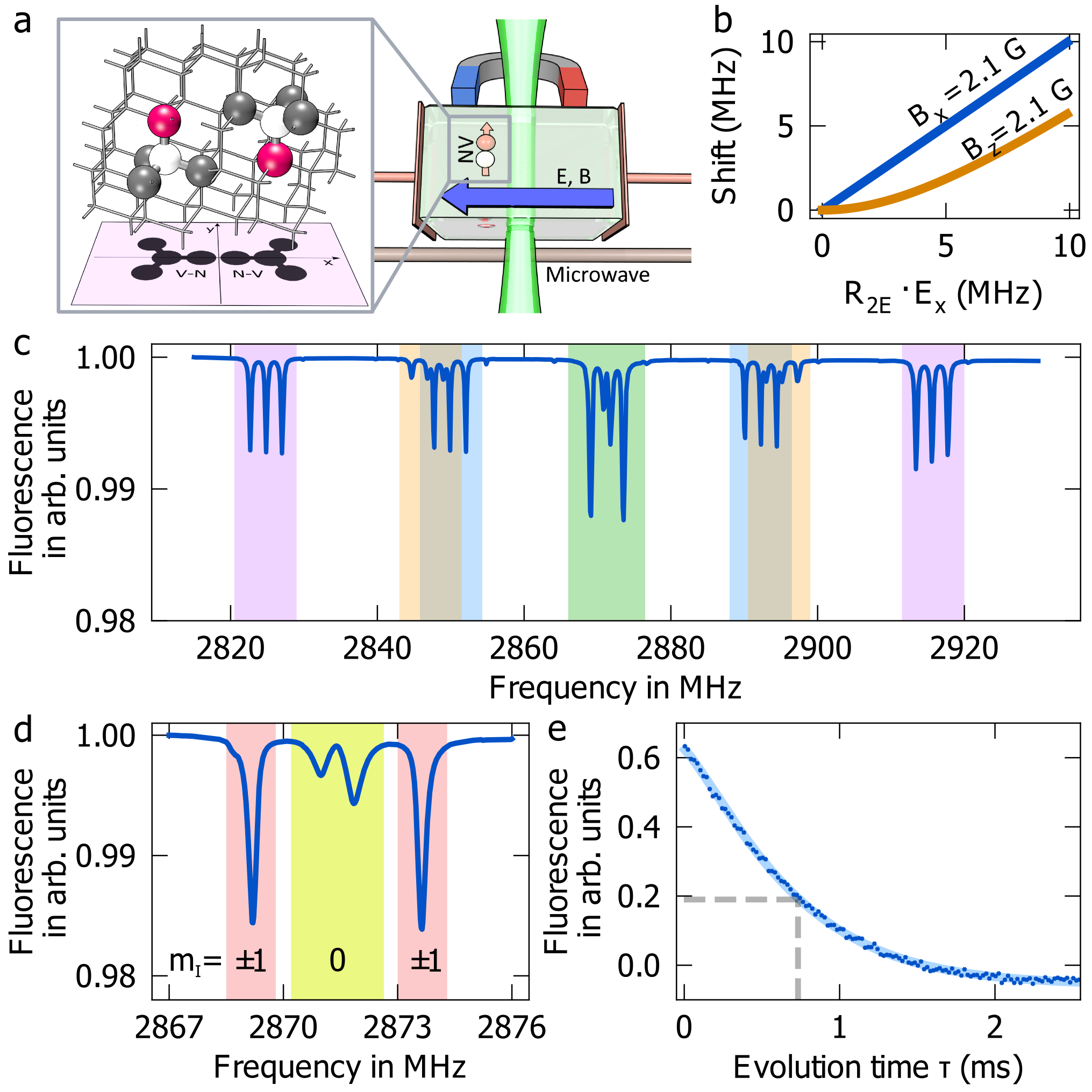}
\caption{Diamond and NV ensemble characteristics. a: A diamond with two (111) surfaces is put into a plate capacitor such that the NVs along the [111]-direction experience a transversal electric field. A transversal magnetic field is applied such that the electric field coupling to the NVs is maximum. The zoom shows the two directions of [111] NVs (N red, V white and adjacent C grey). b: Frequency shift of the spin transitions due to a transversal electric field for a magnetic field of 2.1 G in transversal direction (blue) and in axial direction (orange). c: complete ODMR spectrum of all NV orientations. The dips in the green region belong to the NVs in [111]-direction perpendicular to the surface, on which a transversal magnetic field is applied. d: Detailed spectrum for NVs in [111]-direction, which experience a purely transversal magnetic field. e: Spin-echo measurement on the transition sensitive to electric fields. As these states are much less susceptible to axial magnetic fields, the coherence time of 730 \textmu s is long compared to the states with an axial magnetic field, where $T_2$ is on the order of 200 \textmu s. The $T_1$ relaxation time is 6 ms as shown in the Supporting Information fig. 5(a).}
\label{fig:ensemble}
\end{figure}

%\section{Hamiltonian engineering}
Under standard sensing conditions, the spin of the NV center in the ground state is dominated by the zero-field splitting $D$, followed by the axial magnetic field $B_z$. The resulting eigenstates are thus the eigenstates of the $\hat{S}_z$ operator, noted as $\ket{+1}, \ket{0}$, and $\ket{-1}$. 
The energies of these states do not change in first order with an applied electric field, which can be seen best by rewriting the transversal term of the Stark shift Hamiltonian to
\begin{equation}
H_{\text{Stark}, \perp}=\frac{R_{2E}E_{\perp}}{2}\left(e^{i\phi_E}S_+^2+e^{-i\phi_E}S_-^2\right), 
\label{eq:hshort}
\end{equation}
with the transversal electric field strength $E_{\perp}$ and the angle of the electric field in the $xy$-plane $\phi_E$.
The expectation value of $H_{\text{Stark}, \perp}$ in the basis of $S_z$ is zero:
\begin{equation}
\bra{0}H_{\text{Stark}, \perp}\ket{0}=0, \hspace{0.2cm}\bra{\pm 1}H_{\text{Stark}, \perp}\ket{\pm 1}=0.
\end{equation}
%\begin{equation}
%\left[\hat{H}_{\text{Stark}}, \hat{S}_z\right]&=&\left[E_x\left(\hat{S}_x^2-\hat{S}_y^2\right)-E_y\left(\hat{S}_x\hat{S}_y+\hat{S}_y\hat{S}_x\right), \hat{S}_z\right]\nonumber\\
%&=&0, 
%\end{eqnarray}
Therefore, to measure a linear Stark shift using NVs, the eigenstates of the NV have to be designed to be susceptible to electric fields. Such an eigenbasis can be achieved by three means, applying an electric field strong enough to couple stronger to the NV than the axial magnetic field\cite{Chen2017}, having no magnetic field at all, or applying a transversal magnetic field\cite{Michl2014, Dolde2011, Doherty2014}. The first two options are achievable, but technically difficult. Applying a transversal magnetic field on the order of some millitesla leads to a suppression of the coupling to an axial magnetic field. States dominated by a transversal magnetic field are insensitive towards fluctuating surrounding nuclear spins and exhibit longer coherence times. These states, without an axial magnetic field and a transversal magnetic field much smaller than the zero-field splitting, are
\begin{eqnarray}
\ket{-}&=&\frac{1}{\sqrt{2}}\left(\ket{+1}-e^{2i\phi_B}\ket{-1}\right)\\
\ket{+}&=&\frac{1}{\sqrt{2}}\left(\ket{+1}+e^{2i\phi_B}\ket{-1}\right),
\label{eq:statestransversal}
\end{eqnarray}
where $\phi_B$ is the azimuthal angle of the magnetic field and they show a linear Stark shift, as the expectation values of $H_{\text{Stark}, \perp}$ in this basis become
\begin{eqnarray}
\bra{-}H_{\text{Stark},\perp}\ket{-}&=&R_{2E}E_{\perp}\cos{\left(2\phi_B+\phi_E\right)}\\
\bra{+}H_{\text{Stark},\perp}\ket{+}&=&-R_{2E}E_{\perp}\cos{\left(2\phi_B+\phi_E\right)}.
\end{eqnarray}
This shows that the shift depends significantly on the direction of the transversal electric and magnetic fields relative to each other and to the direction of the three carbon atoms adjacent to the vacancy\cite{Doherty2014}. In this Hamiltonian, $\phi=0$ is defined as being along the projection of one of the vacancy-carbon axes on the transversal plane. %The sign of the proportionality depends on the state to which the transition is driven. 
For the lower transition, the shift is positive at $\phi_B=\phi_E=0$, where both fields are in the direction of a carbon atom. Hence, the shift follows the three-fold symmetry of the NV, as $\phi=\phi_B=\phi_E$ leads to a shift proportional to $\cos{3\phi}$\cite{Michl2014}. 
In an ensemble of NVs, the directions of the NVs are statistically distributed, so along each of the four possible alignment axes for the NV, the order of nitrogen and vacancy is reversed for half of the NVs, as shown in fig. \ref{fig:ensemble}(a). %With the direction of the NV, the direction of the adjacent carbon atoms is also inverted with respect to the NV axis and 
It follows, that for a given applied magnetic and electric transversal field, the frequency shift is positive for half the NVs in the ensemble and negative for the other half of the ensemble. So, only the absolute value of the shift is measurable.
%Measuring a frequency shift well under the linewidth of the transition is not possible as the line does not shift uniformly but splits. It is possible, however, to measure only the absolute value of the shift.
A dense ensemble of NVs also exhibits high strain, which is distributed randomly between the NVs, leading to a broadening of the transition. As the strain Hamiltonian has the same form as the electric field Hamiltonian, a static electric field can not be distinguished from strain and the transitions sensitive to electric fields are inhomogeneously broadened by strain far more than the states dominated by an axial field. \\

%todo:
%\begin{itemize}
%	\item Perform DQ electric field sensing. That yields higher sensitivity and is less prone to temperature fluctuations.
%	\item Try to achieve as narrow as possible ODMR linewidth, maybe via Ramsey spectroscopy. That enables best reconstruction of strain distribution.
%	\item Measure $E$-field uncertainty scaling with highest possible sensitivity (i.e. long $\tau$, no unnecessary dead times, evaluate whole fluorescence, DQ, ...) and for as long as possible (i.e. hours or days). Try to hit baseline, where uncertainty stops decreasing or has a deviation from $1/\sqrt{t}$ scaling.
%\end{itemize}
%

%\section{Results}
%\subsection{Lock-In Detection}

\begin{figure}
\includegraphics[width=\columnwidth]{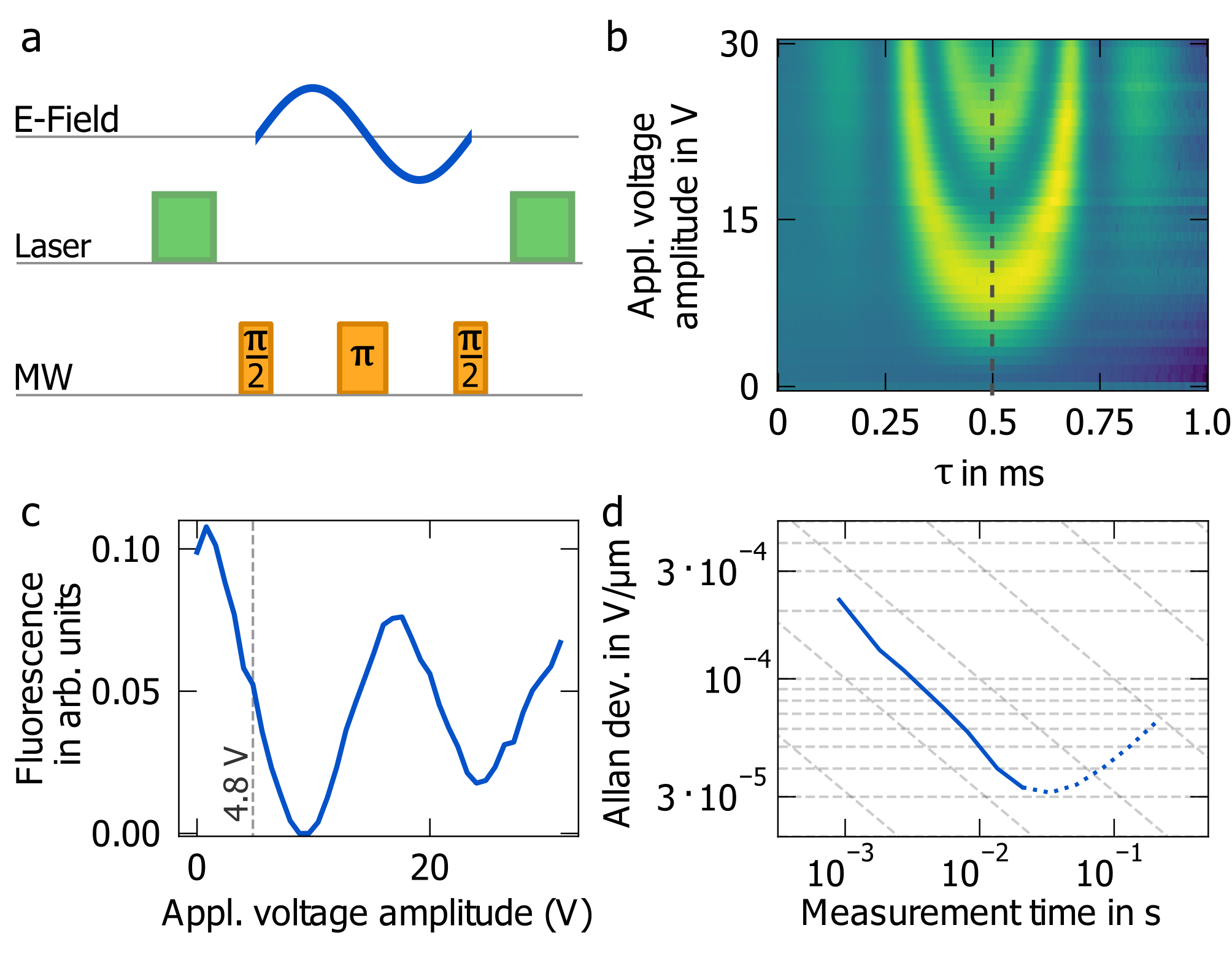}
\caption{Conventional Hahn-Echo measurement. a: An electric field with a frequency of 2 kHz is applied triggered to each measurement sequence. b: measured signal for different evolution times $\tau$ over different voltage amplitudes. c: Signal over applied voltage amplitude at the most sensitive evolution time $\tau=500 \:$\textmu s (dashed line in (b)). The visible decay is due to inhomogeneities of the magnetic and electric field over the sample volume. The sensitivity to electric field changes is dependent on the applied electric field. d: Allan deviation of the electric field signal at the steepest slope of (c), at about 4.8 V amplitude. The sensitivity scales with $1/\sqrt{\text{Hz}}$ only for a short time of some tens of milliseconds (solid vs. dotted line).}
\label{fig:meas1}
\end{figure}

Here we measure the AC electric field of a plate capacitor using a $\approx$ 500 \textmu m large diamond with a 0.9 ppm NV ensemble, see fig. 1. We apply a static magnetic field to taylor the NV spin Hamiltonian as mentioned and use microwave fields to manipulate the spins and perform measurement sequences. The spin states are read out optically through the spin-dependent fluorescence of the NV by applying laser pulses on the ensemble where the resulting PL signal accumulates the response of the NV spins within a sample volume of $\approx$ 500x50x50 \textmu m$^{3}$ in a single sweep.
Conventional metrology involving NVs depends on repetitive measurement schemes where
the signal to be measured is triggered to the measurement sequence and the results
are averaged.\cite{Wolf2015}. The use of a Hahn-Echo measurement, as shown in fig. \ref{fig:meas1}(a), leads to a phase collected by the NV spin during the free evolution time $\tau$ (see fig. \ref{fig:meas1}(b)), which is maximum for $\tau$ being equal to the period of the signal. 
The scheme measures the cosine of the accumulated phase $\Phi_{\text{acc}}$,
\begin{equation}
\Phi_{\text{acc}} = \frac{R_{2E}E}{f}(2\cdot \cos (\pi f \tau)-\cos (2\pi f\tau)-1)
\end{equation}
with the electric field amplitude $E$, the signal frequency $f$ and the evolution time $\tau$. The resulting PL signal response is hence also a cosine which decays for higher electric fields due to electric and magnetic field inhomogenities over the measured NV ensemble. This signal over the applied voltage amplitude is shown in fig. \ref{fig:meas1} (c). Here, the most sensitive point for electric fields is at an applied voltage amplitude of about 4.8 V, where the slope is steepest (see fig. \ref{fig:meas1}(c)). Especially for small electric field signals, the measurement of the sine of the accumulated phase $\Phi_{\text{acc}}$ would be advantageous, yet is not possible here since the sign of the frequency shift is reversed for half of the NVs, as mentioned before. By measuring the noise of the measurement, the Allan deviation can be calculated (see fig. \ref{fig:meas1} (d)) and it can be seen that it scales only up to a few tens of ms, leading to a maximum precision of about 3.1$\cdot 10^{-5}$ V/\textmu m. The maximum phase which can be collected during such a measurement is limited by the electron spin coherence time $T_2$. This time can be prolonged by using advanced dynamical decoupling sequences, e.g. CPMG or XY-8. The sensitivity then scales with $\sqrt{T_{2\rho}}$, when $T_{2\rho}$ is the apparent coherence time under pulsed decoupling. 
Typically $T_{2\rho}$ is significantly longer than $T_2$\cite{reviewFriedemann}. In a quasiclassical picture, the multipulse sequences generate a filter function to suppress noise which is the Fourier transform of the pulse sequence applied. The periodic signal to be measured exactly falls into the transmission window of this filter function, while all noise components outside this transmission window are suppressed. For a dynamical decoupling sequence like CPMG, the width of this spectral window scales with $\tau^{-1}$ when $\tau$ is the overall phase accumulation length and the suppression of noise outside of this spectral window goes with $N^2$, when $N$ is the number of decoupling pulses. Both $\tau$ and $N$ are limited by the coherence time of the defect ($T_{2\rho}$) and the achievable Rabi frequency. As a result the width of the spectral window is essentially limited by the spin properties. All noise components falling within this spectral window will contribute to the signal-to-noise ratio of the measurement. The influence of laser noise on the measurement is also limited by referencing the measurement signal to the simultaneously measured laser power. The Allan deviation plot shown in fig. \ref{fig:meas1}(c) shows that sensitivity only scales as the standard quantum limit up to averaging times of around a few 10 ms. After that low frequency noise components deteriorate further scaling of the precision of the measurement. It is thus of central importance to identify methods which further allow to reduce the spectral width of the filter function to achieve improved suppression particularly of low frequency noise components.
\begin{figure}
\includegraphics[width=\columnwidth]{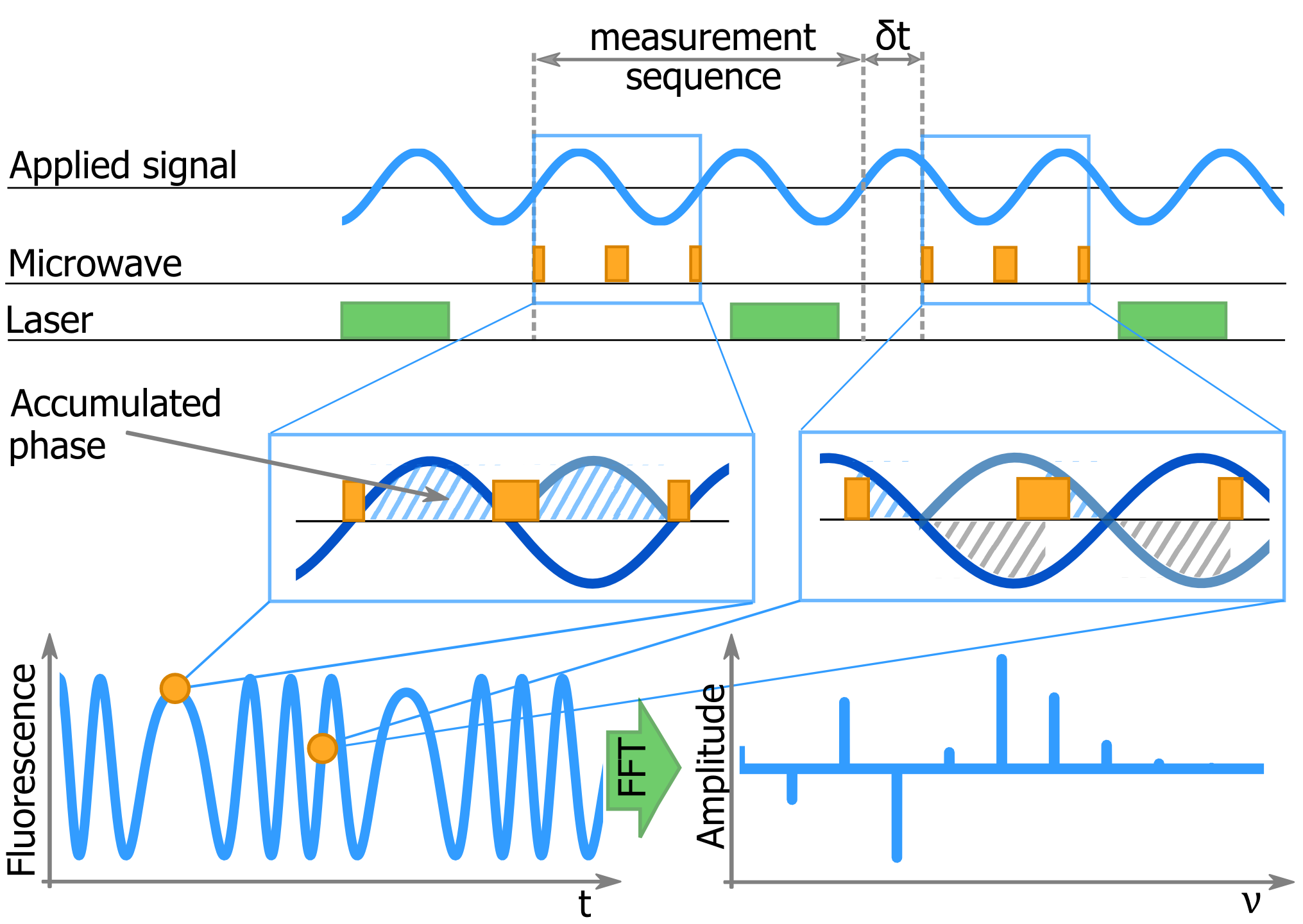}
\caption{Measurement sequence. The single frequency voltage signal is applied continuously and a train of Hahn-Echo measurements with a fixed phase shift $2\pi f\delta t$ with respect to the oscillating signal is conducted. Depending on the relative phase between voltage signal and Hahn-Echo sequence, the phase collected by the NV spins during a measurement changes. This phase becomes maximum ($\Phi_{\text{acc, max}}$) when the signal forms a sine during the measurement (see left box). This leads to a sampling of the applied signal. The fluorescence response of the NV ensemble is the cosine of the collected phase, which goes periodically with the signal. The Fourier transform of the response shows peaks at multiples of $2f$.}
\label{fig:sequence}
\end{figure}
Recently, several works extended the dynamical decoupling scheme for the measurement of periodically oscillating, phase stable signals\cite{Schmitt2017, Boss2017, Glenn2018}. In these schemes the signal acquired by dynamical decoupling sequences is accumulated in a phase coherent manner. By this the filter function of the pulse sequence are further narrowed beyond the value limited by spin relaxation. Ideally, its width is only limited by the total measurement time of the experiment. Previously, this was used for precise measurement of the signal frequencies, as the frequency resolution as well as the bandwidth is not limited by the coherence times of the NV but by the coherence time of the signal. Here, we show the usage of this measurement scheme to measure signal amplitudes with enhanced precision. The main idea is as follows. An AC signal can be measured effectively using a Hahn-Echo measurement, as long as the frequency of the signal is not lower than $1/T_2$. The $\pi$-pulse in the middle of the free evolution time $\tau$ refocuses the spin such that any low frequency fluctuations are averaged out, while an AC signal which changes its sign during the $\pi$-pulse is enhanced.
The phase between measurement and signal is now shifted between subsequent measurements, as the measurement train adds a phase shift of $\delta \phi = 2\pi f \delta t$ towards the signal after each measurement, with $f$ being the signal frequency and $\delta t$ the time added between measurement sequences. The phase collected by the NV is now dependent on the phase of the measured signal compared to the measurement (see fig. \ref{fig:sequence} and Supporting Information):
\begin{equation}
\Phi_{\text{acc}} = \frac{4a}{f}\sin{\left(\phi_0\right)},
\label{eq:phi_liam}
\end{equation}
where $a$ is the amplitude of the frequency shift of the respective NV spin transition caused by the signal and $\phi_0=2\pi f m \delta t$ is the phase of the signal at the start of the $m$-th measurement. The maximum accumulated phase $\Phi_{\text{acc, max}}=4a/f$ is collected when the signal is a sine during the Hahn-Echo measurement, as shown in fig. \ref{fig:sequence} (left). 
As seen in eq. \ref{eq:phi_liam}, the accumulated phase goes linear with the shift $a$ and oscillates with the signal frequency $f$. The accumulated phase $\Phi_{\text{acc}}$ cannot be read out directly, instead either its sine or cosine is retrieved. When the sine of the phase is measured, $s=\sin{(\Phi_{\text{acc}})}$ and the shift $a$ is small, the measured signal can be approximated by $s=\Phi_{\text{acc}}$, resulting in an oscillation at frequency $f$ and an amplitude proportional to $a$\cite{Schmitt2017}. For larger shifts $a$ or measurements of the cosine, this linear approximation does not hold and oscillations at multiples of $f$ appear in the measured signal.
A measurement of the cosine of the phase, $s=\cos{(\Phi_{\text{acc}})}$ as done in this work, is not suitable for weak signals, as it does not show this linear behavior. The measured signal in this case goes with
\begin{equation}
s(t)=\cos{(\Phi_{\text{acc}})}=J_0\left(\frac{4a}{f}\right)+2\sum_{n=1}^{\infty}J_{2n}\left(\frac{4a}{f}\right)\cos{\left(4\pi n f t\right)},
\label{eq:signalcont}
\end{equation}  
where $J_{\alpha}$ is a Bessel function of the first kind. A detailed calculation of the relation is shown in the Supporting Information. An example from such a measurement can be seen in fig. \ref{fig:measurement}(a). 
Eq. \ref{eq:signalcont} shows, that the resulting spectrum exhibits peaks at multiples of $2f$. With a higher signal strength, higher orders of these Fourier components will appear\cite{Kotler2013}. The bandwidth of the measurement with respect to the coupling strength is limited only by the chosen phase difference $2\pi f\delta t$. As this scheme measures phases between $0$ and $\Phi_{\text{acc, max}}$, it is an effective way to circumvent any ambiguity coming from measuring the cosine of the accumulated phase, i.e. the signal for $\Phi_{\text{acc}}+2\pi$ being the same as for $\Phi_{\text{acc}}$\cite{Waldherr2011}. In fig. \ref{fig:measurement}(b) the amplitudes for the different Fourier components are shown over the coupling to the applied electric field. Fig. \ref{fig:measurement} (d) shows the scaling of the sensitivity over measurement time. The scaling is extended by four orders of magnitude in comparison to conventional Hahn-Echo measurement shown in fig. \ref{fig:meas1}(d) as this scheme works in a comparable way to a classical lock-in detection, where all noise away from the signal frequency and its multiples does not affect the measurement. Here, the limit of the achieved sensitivity is the signal coherence, shown by the difference between the scaling with and without applied signal in fig. \ref{fig:measurement}(d). By comparing the amplitudes of different Fourier components, the influence of the contrast can be limited. As the contrast of the measurement can change over time (e.g. by temperature shifts leading to different Rabi frequencies or by microwave power fluctuations), this further improves the scaling of the measurement, which is shown in fig. \ref{fig:measurement}(d) by the difference between the blue and the purple curve.

\begin{figure}
\includegraphics[width=\columnwidth]{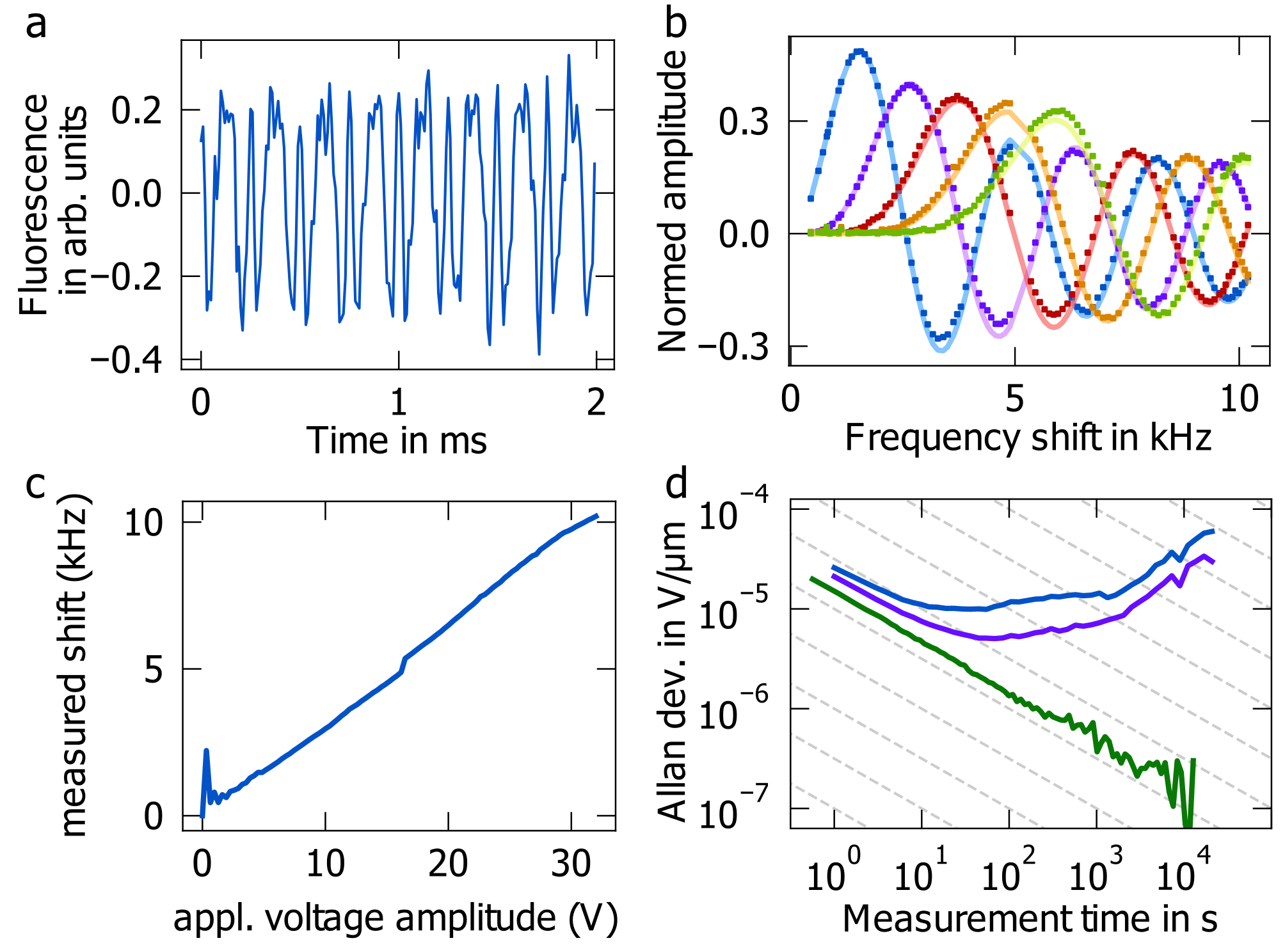}
\caption{Measurement results. a: Exemplary measurement of the electric field signal over the time $t=m\delta t$ between measurements, with the number of measurements $m$ and the wait time $\delta t$ between measurements.  b: Normed amplitudes of peaks within the Fourier transform over the fitted coupling strength %(blue, purple, red, orange, green correspond to Fourier components with 2$f$, 4$f$, 6$f$ and 8$f$, respectively)
. c: Measured coupling strength over applied voltage amplitude. For a small signal, the measurement does not work well, as the accumulated phase is measured via a cosine. The visible jump at 5 V and 16 V is due to the signal generator (see Supporting Information fig. 1). d: Scaling of sensitivity with measurement time. The blue line shows the sensitivity using only the amplitude for the frequency of $2f$, the green line shows the sensitivity calculated by the ratio between the amplitudes at $2f$ and $4f$. It shows a $\sqrt{\text{Hz}}$ behavior for about 100 s, leading to a maximum precision of $\approx 5\cdot10^{-6}$ V/\textmu m. The purple line depicts the sensitivity for noise measured on frequencies which are not a multiple of $f$. As this shows a better sensitivity as well as a longer scaling, the measurement is limited by the signal source.}
\label{fig:measurement}
\end{figure}

%Any field inhomogeneities, which lead to a slightly different phase collected by the different NVs, will lead to a small contribution from neighboring peaks to a peak, thus changing the ratio between the peaks. A detailed calculation of this effect can be found in \cite{Soms}. 
A comparison of the measured coupling strength and the independently verified applied voltage amplitudes lead to a measured shift of the NV transition frequencies of 0.315 kHz/V, which can be recalculated using the gap of the plate capacitor as $d=490 \pm 10$ \textmu m to get the coupling constant $R_{2E}=165 \pm 7$ kHz/V/\textmu m, refining the value given by van Oort\cite{Oort1990} of $R_{2E}=175\pm 30$ kHz/V/\textmu m. A detailed calculation of this coupling constant can be found in the Supporting Information.\\

%\subsection{Double Quantum Transition}

\begin{figure}
\includegraphics[width=\columnwidth]{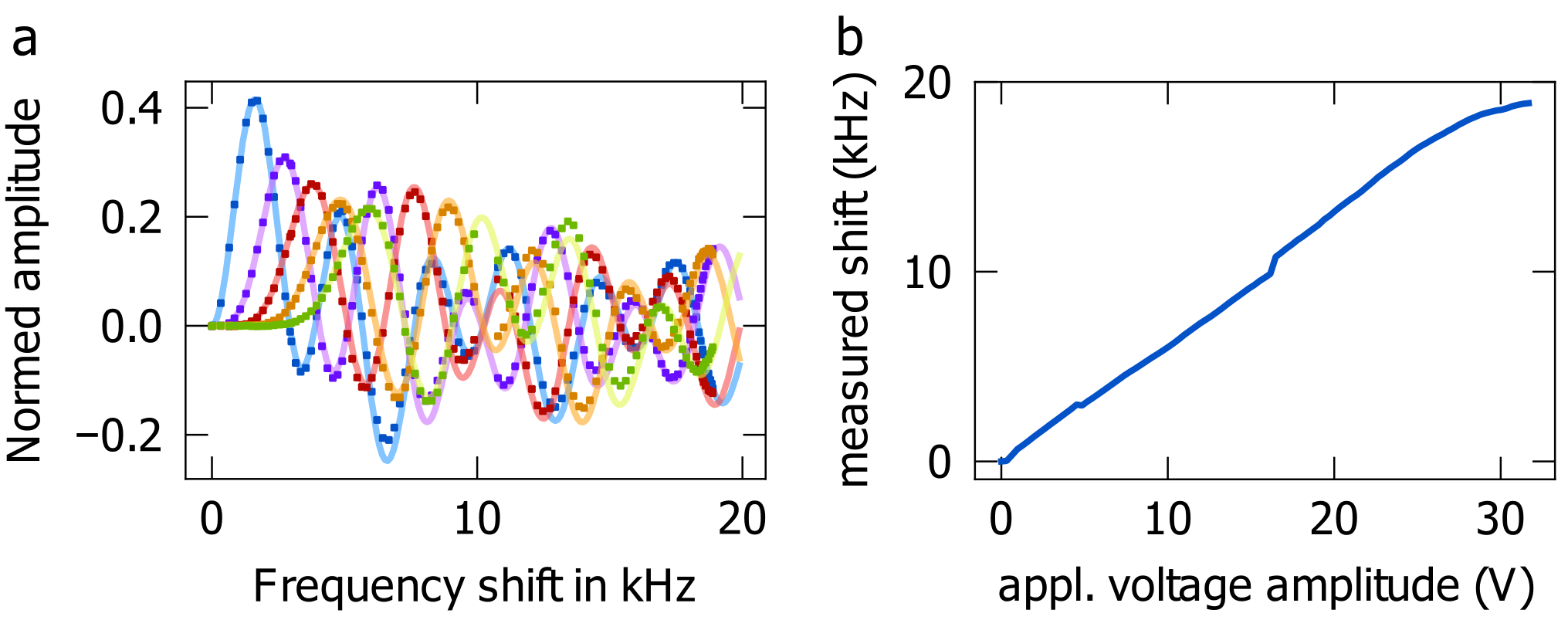}
\caption{Double quantum transition measurement. a: Amplitude of the different orders of Bessel-Functions over the applied electric field amplitude. As the MW pulses do not allow for a high population in the DQT, every amplitude also has a part with half its coupling from the residual SQT. b: with DQT, it is possible to measure lower applied fields than with SQT, as can be seen by comparison to fig. \ref{fig:measurement}(c), the jumps are from the used signal generator (the independently measured voltage over the nominally applied voltage can be seen in the Supporting Information).}
\label{fig:dqt}
\end{figure}

Alternatively to the transition $m_s=0$ to $m_s=\pm 1$, the Stark shift can also be measured in the double quantum transition (DQT) between the two $m_s=\pm 1$ states. For this transition, the frequency shift is doubled, as the two states experience identical shifts but with a different sign. The resulting measurement is shown in fig. \ref{fig:dqt} (a). Due to differences in the transition frequency and microwave coupling in an ensemble, MW pulses will not only excite a double quantum transition, but additionally the single quantum transition (SQT) is visible in the measurement. As the percentage of SQT in the measurement does not depend on the electric field, it is assumed to be constant over all measurements using the same MW pulses. In addition to the advantage of measuring twice the phase, making it possible to measure smaller signals, a DQT measurement also allows for the elimination of temperature influences during a measurement, as both transitions are shifted equally by temperature. The measurement presented here in fig. \ref{fig:dqt} shows the advantage for low signal strength over a SQT measurement. Yet, the electric field sensitivity showed no improvement, meaning either that there is no limiting noise from temperature or the imperfect MW pulses and residual SQT make this advantage void. Here, robust MW pulses and two orthogonal linear polarized MW fields could help in the future to achieve a much higher degree of population in the double quantum transition, making it possible to fully utilize the advantages of a DQT measurement. 

%\section{Conclusion}
In conclusion, we presented precision measurement of electric fields using an ensemble of NV centers. The measurement of AC electric fields much smaller than the intrinsic strain could be shown. 
We achieved a precision of approximately $10^{-5}$ V/\textmu m/$\sqrt{\text{Hz}}$. For this, we adapted continuous measurement schemes and operate them in the non-linear regime. This allows for the measurement of signals over a large bandwidth of signal frequencies and also a large bandwidth of coupling frequencies, akin to a quantum phase estimation algorithm\cite{Waldherr2011}. As we compare the amplitudes at multiples of the known signal frequency, this measurement also functions as a lock-in detection. 
The coupling constant for a coupling between NV and electric field was found to be $R_{2E}=165 \pm 7$ kHz/V/\textmu m, which is in good agreement with the previously reported coupling constant $R_{2E}=175\pm 30$ kHz/V/\textmu m\cite{Oort1990}. The longitudinal coupling constant $R_{3D}$ was too small to be measured, which would match with a coupling constant $R_{3D}=3.5$ kHz/V/\textmu m reported by Van Oort. With measurements at the ground state level anti-crossing, we are able to give an upper limit for the as of yet unmeasured coupling constant $R_{15}<0.047\cdot R_{2E}$ (see Supporting Information). Our measurement scheme might well be extended to other sensing modalities where weak signals are accumulated over extended averaging period including magnetic field and temperature measurements.

%\begin{itemize}
%	\item Directional measurement (oder in die sups?)
%	\item Hahn-Echo measurement over tau
%	\item DQT measurement
%\end{itemize}
%%\subsection{Lock-In detection}
%\begin{itemize}
%	\item Continuously applied signal -> no trigger necessary
%	\item phases measured between 0 and $\phi_{max}$ -> QPEA
%	\item achieved sensitivity
%\end{itemize}
%\subsection{Determination of coupling strength}
%\begin{itemize}
%	\item transversal component: measured to: extract from new structure + DQT
%	\item $S_xS_z$ component: could not be measured -> upper limit (from LAC)
%	\item $E_z$ component: too small -> cosine not possible
%\end{itemize}

%\section{Discussion}

%\appendix

%\section{methods}

%\section{Data availability}
%Data supporting the findings of this study are available within the article and its Methods section and from the corresponding authors upon reasonable request.

\bibliography{bibpaper}

\providecommand{\latin}[1]{#1}
\makeatletter
\providecommand{\doi}
  {\begingroup\let\do\@makeother\dospecials
  \catcode`\{=1 \catcode`\}=2 \doi@aux}
\providecommand{\doi@aux}[1]{\endgroup\texttt{#1}}
\makeatother
\providecommand*\mcitethebibliography{\thebibliography}
\csname @ifundefined\endcsname{endmcitethebibliography}
  {\let\endmcitethebibliography\endthebibliography}{}
\begin{mcitethebibliography}{23}
\providecommand*\natexlab[1]{#1}
\providecommand*\mciteSetBstSublistMode[1]{}
\providecommand*\mciteSetBstMaxWidthForm[2]{}
\providecommand*\mciteBstWouldAddEndPuncttrue
  {\def\EndOfBibitem{\unskip.}}
\providecommand*\mciteBstWouldAddEndPunctfalse
  {\let\EndOfBibitem\relax}
\providecommand*\mciteSetBstMidEndSepPunct[3]{}
\providecommand*\mciteSetBstSublistLabelBeginEnd[3]{}
\providecommand*\EndOfBibitem{}
\mciteSetBstSublistMode{f}
\mciteSetBstMaxWidthForm{subitem}{(\alph{mcitesubitemcount})}
\mciteSetBstSublistLabelBeginEnd
  {\mcitemaxwidthsubitemform\space}
  {\relax}
  {\relax}

\bibitem[Degen \latin{et~al.}(2017)Degen, Reinhard, and
  Cappellaro]{reviewFriedemann}
Degen,~C.~L.; Reinhard,~F.; Cappellaro,~P. {Quantum sensing}. \emph{Reviews of
  Modern Physics} \textbf{2017}, \emph{89}, 1--39\relax
\mciteBstWouldAddEndPuncttrue
\mciteSetBstMidEndSepPunct{\mcitedefaultmidpunct}
{\mcitedefaultendpunct}{\mcitedefaultseppunct}\relax
\EndOfBibitem
\bibitem[Laraoui \latin{et~al.}(2015)Laraoui, Aycock-Rizzo, Gao, Lu, Riedo, and
  Meriles]{Laraoui2015}
Laraoui,~A.; Aycock-Rizzo,~H.; Gao,~Y.; Lu,~X.; Riedo,~E.; Meriles,~C.~A.
  {Imaging thermal conductivity with nanoscale resolution using a scanning spin
  probe}. \emph{Nature Communications} \textbf{2015}, \emph{6}\relax
\mciteBstWouldAddEndPuncttrue
\mciteSetBstMidEndSepPunct{\mcitedefaultmidpunct}
{\mcitedefaultendpunct}{\mcitedefaultseppunct}\relax
\EndOfBibitem
\bibitem[Lee \latin{et~al.}(2008)Lee, Zhu, and Seshia]{Lee2008}
Lee,~J.; Zhu,~Y.; Seshia,~A. Room temperature electrometry with SUB-10 electron
  charge resolution. \emph{Journal of Micromechanics and Microengineering}
  \textbf{2008}, \emph{18}, 025033\relax
\mciteBstWouldAddEndPuncttrue
\mciteSetBstMidEndSepPunct{\mcitedefaultmidpunct}
{\mcitedefaultendpunct}{\mcitedefaultseppunct}\relax
\EndOfBibitem
\bibitem[Vincent \latin{et~al.}(2004)Vincent, Narayan, Pettersson, Willander,
  Jeppson, and Bengtsson]{Vincent2004}
Vincent,~J.~K.; Narayan,~V.; Pettersson,~H.; Willander,~M.; Jeppson,~K.;
  Bengtsson,~L. Theory of a room-temperature silicon quantum dot device as a
  sensitive electrometer. \emph{Journal of Applied Physics} \textbf{2004},
  \emph{95}, 323--326\relax
\mciteBstWouldAddEndPuncttrue
\mciteSetBstMidEndSepPunct{\mcitedefaultmidpunct}
{\mcitedefaultendpunct}{\mcitedefaultseppunct}\relax
\EndOfBibitem
\bibitem[Neumann \latin{et~al.}(2013)Neumann, Volk, Engels, and
  Stampfer]{CNeumann2013}
Neumann,~C.; Volk,~C.; Engels,~S.; Stampfer,~C. Graphene-based charge sensors.
  \emph{Nanotechnology} \textbf{2013}, \emph{24}, 444001\relax
\mciteBstWouldAddEndPuncttrue
\mciteSetBstMidEndSepPunct{\mcitedefaultmidpunct}
{\mcitedefaultendpunct}{\mcitedefaultseppunct}\relax
\EndOfBibitem
\bibitem[Osterwalder and Merkt(1999)Osterwalder, and Merkt]{Osterwalder1999}
Osterwalder,~A.; Merkt,~F. {Using high rydberg states as electric field
  sensors}. \emph{Physical Review Letters} \textbf{1999}, \emph{82},
  1831--1834\relax
\mciteBstWouldAddEndPuncttrue
\mciteSetBstMidEndSepPunct{\mcitedefaultmidpunct}
{\mcitedefaultendpunct}{\mcitedefaultseppunct}\relax
\EndOfBibitem
\bibitem[Sedlacek \latin{et~al.}(2012)Sedlacek, Schwettmann, K{\"{u}}bler,
  L{\"{o}}w, Pfau, and Shaffer]{Sedlacek2012}
Sedlacek,~J.~A.; Schwettmann,~A.; K{\"{u}}bler,~H.; L{\"{o}}w,~R.; Pfau,~T.;
  Shaffer,~J.~P. {Microwave electrometry with Rydberg atoms in a vapour cell
  using bright atomic resonances}. \emph{Nature Physics} \textbf{2012},
  \emph{8}, 819--824\relax
\mciteBstWouldAddEndPuncttrue
\mciteSetBstMidEndSepPunct{\mcitedefaultmidpunct}
{\mcitedefaultendpunct}{\mcitedefaultseppunct}\relax
\EndOfBibitem
\bibitem[Mims(1976)]{Mims}
Mims,~W. \emph{The linear electric field effect in paramagnetic resonance};
  Clarendon Press, 1976\relax
\mciteBstWouldAddEndPuncttrue
\mciteSetBstMidEndSepPunct{\mcitedefaultmidpunct}
{\mcitedefaultendpunct}{\mcitedefaultseppunct}\relax
\EndOfBibitem
\bibitem[Dolde \latin{et~al.}(2011)Dolde, Fedder, Doherty, N{\"{o}}bauer,
  Rempp, Balasubramanian, Wolf, Reinhard, Hollenberg, Jelezko, and
  Wrachtrup]{Dolde2011}
Dolde,~F.; Fedder,~H.; Doherty,~M.~W.; N{\"{o}}bauer,~T.; Rempp,~F.;
  Balasubramanian,~G.; Wolf,~T.; Reinhard,~F.; Hollenberg,~L.~C.; Jelezko,~F.;
  Wrachtrup,~J. {Electric-field sensing using single diamond spins}.
  \emph{Nature Physics} \textbf{2011}, \emph{7}, 459--463\relax
\mciteBstWouldAddEndPuncttrue
\mciteSetBstMidEndSepPunct{\mcitedefaultmidpunct}
{\mcitedefaultendpunct}{\mcitedefaultseppunct}\relax
\EndOfBibitem
\bibitem[Kato \latin{et~al.}(2003)Kato, Myers, Driscoll, Gossard, Levy, and
  Awschalom]{Kato2003}
Kato,~Y.; Myers,~R.~C.; Driscoll,~D.~C.; Gossard,~A.~C.; Levy,~J.;
  Awschalom,~D.~D. Gigahertz Electron Spin Manipulation Using
  Voltage-Controlled g-Tensor Modulation. \emph{Science} \textbf{2003},
  \emph{299}, 1201--1204\relax
\mciteBstWouldAddEndPuncttrue
\mciteSetBstMidEndSepPunct{\mcitedefaultmidpunct}
{\mcitedefaultendpunct}{\mcitedefaultseppunct}\relax
\EndOfBibitem
\bibitem[Iwasaki \latin{et~al.}(2017)Iwasaki, Naruki, Tahara, Makino, Kato,
  Ogura, Takeuchi, Yamasaki, and Hatano]{Iwasaki2017}
Iwasaki,~T.; Naruki,~W.; Tahara,~K.; Makino,~T.; Kato,~H.; Ogura,~M.;
  Takeuchi,~D.; Yamasaki,~S.; Hatano,~M. {Direct Nanoscale Sensing of the
  Internal Electric Field in Operating Semiconductor Devices Using Single
  Electron Spins}. \emph{ACS Nano} \textbf{2017}, \emph{11}, 1238--1245\relax
\mciteBstWouldAddEndPuncttrue
\mciteSetBstMidEndSepPunct{\mcitedefaultmidpunct}
{\mcitedefaultendpunct}{\mcitedefaultseppunct}\relax
\EndOfBibitem
\bibitem[Wolfowicz \latin{et~al.}(2018)Wolfowicz, Whiteley, and
  Awschalom]{Wolfowicz2018}
Wolfowicz,~G.; Whiteley,~S.~J.; Awschalom,~D.~D. {Electrometry by optical
  charge conversion of deep defects in 4H-SiC}. \emph{Proceedings of the
  National Academy of Sciences} \textbf{2018}, \emph{115}, 7879--7883\relax
\mciteBstWouldAddEndPuncttrue
\mciteSetBstMidEndSepPunct{\mcitedefaultmidpunct}
{\mcitedefaultendpunct}{\mcitedefaultseppunct}\relax
\EndOfBibitem
\bibitem[Chen \latin{et~al.}(2017)Chen, Clevenson, Johnson, Pham, Englund,
  Hemmer, and Braje]{Chen2017}
Chen,~E.~H.; Clevenson,~H.~A.; Johnson,~K.~A.; Pham,~L.~M.; Englund,~D.~R.;
  Hemmer,~P.~R.; Braje,~D.~A. High-sensitivity spin-based electrometry with an
  ensemble of nitrogen-vacancy centers in diamond. \emph{Phys. Rev. A}
  \textbf{2017}, \emph{95}, 053417\relax
\mciteBstWouldAddEndPuncttrue
\mciteSetBstMidEndSepPunct{\mcitedefaultmidpunct}
{\mcitedefaultendpunct}{\mcitedefaultseppunct}\relax
\EndOfBibitem
\bibitem[Oort and Glasbeek(1990)Oort, and Glasbeek]{Oort1990}
Oort,~E.~V.; Glasbeek,~M. {Electric-Field-Induced Modulation of Spin Echoes of
  N-V Centers in Diamond}. \emph{Chemical Physics Letters} \textbf{1990},
  \emph{168}, 529--532\relax
\mciteBstWouldAddEndPuncttrue
\mciteSetBstMidEndSepPunct{\mcitedefaultmidpunct}
{\mcitedefaultendpunct}{\mcitedefaultseppunct}\relax
\EndOfBibitem
\bibitem[Michl \latin{et~al.}(2014)Michl, Teraji, Zaiser, Jakobi, Waldherr,
  Dolde, Neumann, Doherty, Manson, Isoya, and Wrachtrup]{Michl2014}
Michl,~J.; Teraji,~T.; Zaiser,~S.; Jakobi,~I.; Waldherr,~G.; Dolde,~F.;
  Neumann,~P.; Doherty,~M.~W.; Manson,~N.~B.; Isoya,~J.; Wrachtrup,~J. {Perfect
  alignment and preferential orientation of nitrogen-vacancy centers during
  chemical vapor deposition diamond growth on (111) surfaces}. \emph{Applied
  Physics Letters} \textbf{2014}, \emph{104}, 102407\relax
\mciteBstWouldAddEndPuncttrue
\mciteSetBstMidEndSepPunct{\mcitedefaultmidpunct}
{\mcitedefaultendpunct}{\mcitedefaultseppunct}\relax
\EndOfBibitem
\bibitem[Doherty \latin{et~al.}(2014)Doherty, Michl, Dolde, Jakobi, Neumann,
  Manson, and Wrachtrup]{Doherty2014}
Doherty,~M.~W.; Michl,~J.; Dolde,~F.; Jakobi,~I.; Neumann,~P.; Manson,~N.~B.;
  Wrachtrup,~J. {Measuring the defect structure orientation of a single NV-
  centre in diamond}. \emph{New Journal of Physics} \textbf{2014},
  \emph{16}\relax
\mciteBstWouldAddEndPuncttrue
\mciteSetBstMidEndSepPunct{\mcitedefaultmidpunct}
{\mcitedefaultendpunct}{\mcitedefaultseppunct}\relax
\EndOfBibitem
\bibitem[Wolf \latin{et~al.}(2015)Wolf, Neumann, Nakamura, Sumiya, Ohshima,
  Isoya, and Wrachtrup]{Wolf2015}
Wolf,~T.; Neumann,~P.; Nakamura,~K.; Sumiya,~H.; Ohshima,~T.; Isoya,~J.;
  Wrachtrup,~J. {Subpicotesla diamond magnetometry}. \emph{Physical Review X}
  \textbf{2015}, \emph{5}, 1--10\relax
\mciteBstWouldAddEndPuncttrue
\mciteSetBstMidEndSepPunct{\mcitedefaultmidpunct}
{\mcitedefaultendpunct}{\mcitedefaultseppunct}\relax
\EndOfBibitem
\bibitem[Schmitt \latin{et~al.}(2017)Schmitt, Gefen, St{\"u}rner, Unden, Wolff,
  M{\"u}ller, Scheuer, Naydenov, Markham, Pezzagna, Meijer, Schwarz, Plenio,
  Retzker, McGuinness, and Jelezko]{Schmitt2017}
Schmitt,~S. \latin{et~al.}  Submillihertz magnetic spectroscopy performed with
  a nanoscale quantum sensor. \emph{Science} \textbf{2017}, \emph{356},
  832--837\relax
\mciteBstWouldAddEndPuncttrue
\mciteSetBstMidEndSepPunct{\mcitedefaultmidpunct}
{\mcitedefaultendpunct}{\mcitedefaultseppunct}\relax
\EndOfBibitem
\bibitem[Boss \latin{et~al.}(2017)Boss, Cujia, Zopes, and Degen]{Boss2017}
Boss,~J.~M.; Cujia,~K.~S.; Zopes,~J.; Degen,~C.~L. Quantum sensing with
  arbitrary frequency resolution. \emph{Science} \textbf{2017}, \emph{356},
  837--840\relax
\mciteBstWouldAddEndPuncttrue
\mciteSetBstMidEndSepPunct{\mcitedefaultmidpunct}
{\mcitedefaultendpunct}{\mcitedefaultseppunct}\relax
\EndOfBibitem
\bibitem[Glenn \latin{et~al.}(2018)Glenn, Bucher, Lee, Lukin, Park, and
  Walsworth]{Glenn2018}
Glenn,~D.~R.; Bucher,~D.~B.; Lee,~J.; Lukin,~M.~D.; Park,~H.; Walsworth,~R.~L.
  {High-resolution magnetic resonance spectroscopy using a Solid-State spin
  sensor}. \emph{Nature} \textbf{2018}, \emph{555}, 351--354\relax
\mciteBstWouldAddEndPuncttrue
\mciteSetBstMidEndSepPunct{\mcitedefaultmidpunct}
{\mcitedefaultendpunct}{\mcitedefaultseppunct}\relax
\EndOfBibitem
\bibitem[Kotler \latin{et~al.}(2013)Kotler, Akerman, Glickman, and
  Ozeri]{Kotler2013}
Kotler,~S.; Akerman,~N.; Glickman,~Y.; Ozeri,~R. {Nonlinear single-spin
  spectrum analyzer}. \emph{Physical Review Letters} \textbf{2013}, \emph{110},
  1--5\relax
\mciteBstWouldAddEndPuncttrue
\mciteSetBstMidEndSepPunct{\mcitedefaultmidpunct}
{\mcitedefaultendpunct}{\mcitedefaultseppunct}\relax
\EndOfBibitem
\bibitem[Waldherr \latin{et~al.}(2012)Waldherr, Beck, Neumann, Said, Nitsche,
  Markham, Twitchen, Twamley, Jelezko, and Wrachtrup]{Waldherr2011}
Waldherr,~G.; Beck,~J.; Neumann,~P.; Said,~R.~S.; Nitsche,~M.; Markham,~M.~L.;
  Twitchen,~D.~J.; Twamley,~J.; Jelezko,~F.; Wrachtrup,~J. {High-dynamic-range
  magnetometry with a single nuclear spin in diamond}. \emph{Nature
  Nanotechnology} \textbf{2012}, \emph{7}, 105--108\relax
\mciteBstWouldAddEndPuncttrue
\mciteSetBstMidEndSepPunct{\mcitedefaultmidpunct}
{\mcitedefaultendpunct}{\mcitedefaultseppunct}\relax
\EndOfBibitem
\end{mcitethebibliography}

\section{Acknowledgments}
%\begin{acknowledgments}
We want to thank Ingmar Jakobi and Tetyana Shalomayeva for their support with the experiments. Furthermore, we acknowledge the valuable input by Marcus Doherty, especially with regards to the measurement of the $R_{15}$ constant.
We acknowledge the support of the European Commission Marie Curie ETN “QuSCo” (GA N°765267),  and the German science foundation (SPP 1601). This work was also supported by ERC grant SMel, the Max Planck Society and the Humboldt Foundation and the EU - FET Flagship on Quantum Technologies through the Project ASTERIQS. In addition, we acknowlegde the support of the Japan Society of the Promotion of Science (JSPS) KAKENHI (No. 17H02751).
%\end{acknowledgments}

\section{Author contributions}
JM, JS, AD and PN carried out the experiments. AB and AZ gave technical support for the experiments. KN, HS, SO and JI synthesized and characterized the diamond sample. JW supervised the research work. All authors contributed to preparation of the manuscript.

\section{Competing financial interests}
All authors declare not to have any competing financial interests.

\end{document}